\documentclass[11pt,a4paper]{article}
\usepackage[utf8]{inputenc}
\usepackage[T1]{fontenc}
\usepackage[margin=1in]{geometry}
\usepackage{graphicx}
\usepackage{booktabs}
\usepackage{longtable}
\usepackage{array}
\usepackage{calc}
\usepackage{caption}
\usepackage{amsmath}
\usepackage{amssymb}
\usepackage{hyperref}
\usepackage{url}
\usepackage{xcolor}
\hypersetup{colorlinks=true, linkcolor=black, citecolor=blue, urlcolor=blue}

\title{Task-Level AI Readiness Assessment for Business Process Management: \\
The T-IPO Model and LARA Matrix in Financial-Services IT Operations}

\author{
  Mingjun Li \quad Xiaojun Ye\thanks{Corresponding author.} \\
  \small Tsinghua University, Beijing, China \\
  \small \texttt{ li-mj24@mails.tsinghua.edu.cn} \quad \texttt{yexj@tsinghua.edu.cn}
}

\date{April 16, 2026}

\begin{document}

\maketitle

\begin{abstract}
Which tasks inside an enterprise workflow can a large-language-model agent reliably handle, and under what conditions? Most business process modeling frameworks still answer this at the activity level, even though a single activity can bundle work of radically different difficulty. This paper takes the analysis a step smaller. We describe two design artifacts developed in a financial-services IT setting: T-IPO, which represents each task as an eight-element tuple, and LARA (LLM Agent Readiness Assessment), a five-dimension rubric that scores a task's readiness for agent substitution. Compliance Sensitivity carries 1.5$\times$ weight, a value we fixed through a three-round Delphi study and cross-checked with AHP. The rubric produces four levels, L1 to L4, and applies a floor rule so that a task with maximum compliance load cannot be classified below L3 no matter what the other scores say. Both artifacts sit inside a larger methodology (PARTIS) that we map onto BWW ontology in Section 3. We evaluate the instruments across 127 tasks. Inter-rater agreement reaches Fleiss' $\kappa$ = 0.80; a replication at three further institutions returns $\kappa$ = 0.73. A controlled comparison against activity-level assessment suggests, though does not prove, an improvement in predictive utility at the task level. Pilot deployment of 120 task instances confirms that auto-completion decays monotonically from 95\% at L1 through about 70\% at L2 to about 40\% at L3. Exploratory factor analysis points to a two-factor structure: task readiness seems to be determined jointly by cognitive-execution complexity and governance-compliance intensity. We close with a recalibration procedure (LARA-TCA) so the rubric can keep pace with evolving LLM capabilities.
\end{abstract}


\medskip

\section{Introduction}

Brynjolfsson and Mitchell [1] observe in \emph{Science} that ``most occupations are heterogeneous bundles of tasks.`` The rapid maturation of LLM agents---from GPT-4`s advanced reasoning [2] and Eloundou et al.`s [3] task-level impact analysis in \emph{Science} to production-grade platforms---has rendered this insight operationally urgent: \emph{which tasks can reliably be delegated to an LLM agent, under what governance constraints, and in what sequence?}\footnote{A shortened workshop version (10--14 pages) has been submitted to the BPM 2026 Workshop (deadline 3 June 2026); the camera-ready proceedings are expected in Springer LNBIP (September 2026). This preprint is the complete version, intended as the citation anchor for companion papers.}

Current BPM frameworks stumble on this question in several ways. Granularity is the first obstacle: BPMN 2.0 [4] treats the activity as the leaf node, but Dell'Acqua et al. [5] describe a "jagged technological frontier" within single activities, where some subtasks excel with LLMs while adjacent ones fail outright. Assessment is a second difficulty. Frey and Osborne [6] work at the occupation level; Autor et al. [7] offer a binary routine/non-routine classification that LLM emergent capabilities have largely invalidated. In regulated industries, a third issue becomes acute: compliance constraints are hard constraints, but BPMN's text-annotation mechanism lacks the structural expressiveness to encode them. Finally, enterprises need deployment pathways that respect how adoption actually diffuses through an organization [8, 9], yet no theory-grounded phased framework currently exists for agent deployment. We address the first three issues in this paper through two artifacts validated in IT operations; the pathway question is taken up in a companion 4I-Path paper. Generalization beyond IT operations is future work.

The two artifacts follow the DSR paradigm [10, 11]. Following Peffers et al.`s [12] DSRM stages: (i) problem identification through practitioner observation; (ii) objectives defined as task-level formalization and quantitative assessment; (iii) artifacts designed through iterative expert review; (iv) demonstration across 127 tasks; (v) multi-method evaluation; (vi) communication via this paper. The knowledge contribution targets Gregor and Hevner`s [13] Level 2 (nascent design theory). Both artifacts sit inside the larger PARTIS methodology (Process--Activity--Role--Task--Institution--Standard), whose six dimensions form dual cycles of execution flow (P$\rightarrow$A$\rightarrow$R$\rightarrow$T) and governance flow (T$\rightarrow$I$\rightarrow$S$\rightarrow$P), detailed in Section 3. The fourth gap (pathway) is addressed by the companion 4I-Path framework, targeting ICIS 2026.

Our empirical work was carried out entirely inside financial-services IT operations. On the primary 127-task dataset, Fleiss' $\kappa$ reached 0.80 (Gwet's AC1 = 0.77 [14, 15]). A second wave of replication at three further institutions returned $\kappa$ = 0.73. A controlled head-to-head against activity-level assessment gave preliminary, though not definitive, evidence of improved predictive utility at the task level. Pilot deployment of 120 task instances showed a monotonic decay in auto-completion from 95\% at L1 (114/120 instances) through about 70\% at L2 to about 40\% at L3. Exploratory factor analysis pulled out a two-factor structure structurally consistent with the architecture's two cycles. These results primarily establish inter-rater reliability, pilot utility, and preliminary predictive usefulness in the target context, not universal superiority across domains or assessment paradigms.

The remainder is organized as follows. Section 2 reviews related work. Section 3 introduces the PARTIS framework. Section 4 presents T-IPO. Section 5 details LARA. Section 6 reports empirical validation. Section 7 discusses contributions and threats. Section 8 concludes.

\section{Related Work}

\subsection{Task-Level Automation Assessment}

Autor, Levy, and Murnane [7] classify tasks along routine/non-routine and cognitive/manual axes. LLM emergent abilities [2, 16] have breached this boundary. Frey and Osborne [6] estimate that 47\% of U.S. occupations face high automation risk at the occupation granularity; Arntz et al. [17], analyzing at the task level, obtain 9\%---underscoring sensitivity to the unit of analysis. Noy and Zhang [38] provide controlled experimental evidence in Science that ChatGPT reduces individual task completion time by 40\% and increases output quality, confirming that task-level productivity effects are both real and measurable. Eloundou et al. [3] provide the closest methodological precedent to LARA: their GPT-exposure rubric evaluates task-level LLM susceptibility across O*NET occupations. However, their rubric stops short in three respects: it has no governance-dimension integration, no formal task decomposition below the O*NET level, and no deployment-mode guidance that would link scores to organizational decisions. Dell`Acqua et al. [5] demonstrate the ``jagged technological frontier``: within the same occupation, some tasks see dramatic AI productivity gains while adjacent tasks see none. LARA responds by operating at the task granularity with a five-dimensional scheme anchored in Bloom`s revised taxonomy [18].

\subsection{BPM and AI Integration}

Dumas et al. [19] publish a research manifesto calling for ``AI-augmented BPM systems.`` Kaltenpoth et al. [20] demonstrate at BPM 2025 that integrating LLM agents with process rules reduces failure rates from 16\% to 1\%. However, neither provides a task-level assessment framework. BPMN 2.0 [4] treats activities as atomic leaf nodes without mandatory IPO decomposition [21]. Reichert and Weber [50] address process flexibility but not agent-substitutability assessment. Process mining [22] can discover as-is models but does not evaluate agent substitutability. Hughes et al. [23] survey agentic systems but do not address structured task-level readiness assessment. In the compliance-aware BPM literature, approaches such as compliance-by-design [39] embed regulatory constraints into process models at design time---PARTIS`s governance flow (T$\rightarrow$I$\rightarrow$S$\rightarrow$P) extends this principle from compliance checking to AI-readiness assessment. Decision modeling via DMN [40] complements PARTIS by formalizing decision logic at the activity level; T-IPO pushes this to task-level Logic elements with a cognitive classification grounded in Bloom's taxonomy.

\subsection{Task Analysis Methods}

Classical methods---HTA [24], GOMS [25], CTA [26]---share with T-IPO the idea of decomposing complex work. Three differences are critical. First, purpose: HTA targets human-factors risk; T-IPO targets AI substitutability and agent-prompt generation. Second, termination: HTA uses the P$\times$C risk rule (safety-oriented); T-IPO uses LARA-assessability (evaluation-oriented)---a qualitatively different stopping condition. Third, output: T-IPO produces formal eight-tuples directly mappable to prompt specifications. T-IPO occupies a vacant niche: a formal task analysis method for AI substitutability with governance integration.

\subsection{AI Governance}

The EU AI Act [27], NIST AI RMF [28], and Floridi et al.`s AI4People [29] establish governance principles but do not prescribe structural embedding within process models. Wang et al.`s AgentSpec [47] provides runtime enforcement for LLM agents but operates at the execution level without task-level readiness assessment. Scott`s [30] three-pillar framework (extending North`s [46] institutional economics) provides the theoretical lens. PARTIS operationalizes this through its Institution and Standard dimensions, corresponding to Scott`s regulative and normative pillars.

\subsection{Positioning: Systematic Comparison}

Table 1 positions T-IPO + LARA against major alternative approaches along four dimensions critical for LLM-agent deployment decisions.

\begin{table}[!htbp]
\centering
\small
\renewcommand{\arraystretch}{1.1}
\begin{tabular}{@{}>{\raggedright\arraybackslash}p{0.1672\linewidth}
  >{\raggedright\arraybackslash}p{0.1770\linewidth}
  >{\raggedright\arraybackslash}p{0.1770\linewidth}
  >{\raggedright\arraybackslash}p{0.1770\linewidth}
  >{\raggedright\arraybackslash}p{0.2217\linewidth}@{}}
\toprule
\textbf{Approach} & \textbf{Modeling granularity} & \textbf{Governance embedding} & \textbf{Prompt mappability} & \textbf{Deployment guidance} \\
BPMN activity-level & Activity (coarse) & Text annotation only & None & None \\
HTA / CTA / GOMS & Operation / cognitive & None & None & None \\
Eloundou et al. GPT-exposure & O*NET task & None & None & Binary (exposed / not) \\
Compliance-aware BPM & Activity & Structural (rules) & None & Compliance check \\
\textbf{T-IPO + LARA} & \textbf{Task (8-tuple)} & \textbf{Structural (I+S dims)} & \textbf{Direct (4-section)} & \textbf{4-level + mode} \\
\bottomrule
\end{tabular}
\caption{Systematic positioning of T-IPO + LARA against alternative approaches.}
\label{tab:1}
\end{table}

Conceptually, T-IPO contributes to the BPM discourse on process abstraction by introducing a sub-activity decomposition layer with formal termination criteria (centered on LARA-assessability, not safety risk). This extends the granularity spectrum that currently runs from process to activity. LARA contributes to compliance-aware BPM [39] by shifting from binary compliance checking (``compliant vs. non-compliant``) to graded readiness assessment (``how much governance oversight does this task require under agent execution?``). The T-IPO--to--prompt mapping connects to the resource-aware BPM literature by treating the LLM agent as a typed resource whose assignment is conditioned on task-level attributes---analogous to role-based resource allocation but with capability-based selection criteria.

\subsection{From RPA to Cognitive Automation}

Robotic Process Automation (RPA) [41] automates deterministic, rule-based tasks via UI-level screen scraping. Van der Aalst et al. [41] position RPA as ``surface-level`` automation that neither understands nor reasons about the underlying process logic. McKinsey [49] estimates that generative AI could automate 60--70\% of employee time across knowledge work---but this aggregate masks enormous task-level variation. Acemoglu and Restrepo [42] provide the theoretical grounding: their task-based framework shows that automation displaces labor from existing tasks while simultaneously creating new tasks---LLM agents accelerate both dynamics, making task-level assessment (rather than occupation-level prediction) the appropriate unit of analysis. The emergence of LLM agents represents a qualitative shift: from deterministic rule execution to probabilistic reasoning under uncertainty. This shift invalidates RPA`s binary classification of tasks as ``automatable`` or ``not automatable``---LLM agents can attempt tasks previously considered non-automatable, but with varying success rates and governance implications. PARTIS`s four-level LARA classification (L1--L4) captures this continuous spectrum, while the T-IPO`s Determinism attribute (Deterministic/Probabilistic/Heuristic) explicitly marks where the automation boundary has shifted from rule execution to probabilistic generation.

\subsection{Ontological Grounding: BWW Representation}

The Bunge--Wand--Weber (BWW) ontology [32] provides a formal criterion for evaluating modeling language completeness: a modeling grammar should represent all ontological constructs relevant to its application domain. PARTIS maps to BWW as follows: Process and Activity map to BWW`s ``thing`` construct; Task maps to ``property`` (an attribute of Activity observable through its IPO structure); Role maps to ``kind`` (a classification of execution agents); Institution and Standard map to ``law`` (constraints governing thing behavior); and LARA Score maps to ``property value`` (an assessed attribute of the Task at a given time). The principal BWW gap is ``event``---PARTIS does not model temporal events as first-class elements, delegating event handling to the BPMN layer. This deliberate incompleteness is an architectural decision, not an oversight: it positions PARTIS as an extension grammar that sits alongside BPMN/CMMN/DMN, not a standalone replacement for them. The BWW analysis thus reveals that PARTIS`s design space is precisely the space that BPMN leaves uncovered---task-level structure, governance constraints, and readiness assessment---while BPMN`s design space covers what PARTIS deliberately omits---event semantics, control flow, and execution choreography. This layering is a deliberate design choice, not an oversight.

\section{The PARTIS Framework}

PARTIS (Process--Activity--Role--Task--Institution--Standard) is a six-dimensional BPM methodology for systematic LLM-agent substitutability assessment, designed following Zachman`s [31] multi-perspective principle and grounded in the BWW ontology [32, 45] for representational completeness. The six dimensions are organized as a hexagonal architecture with two complementary cycles (Fig. 1):

\begin{figure}[!htbp]
\centering
\includegraphics[width=0.689\linewidth,keepaspectratio]{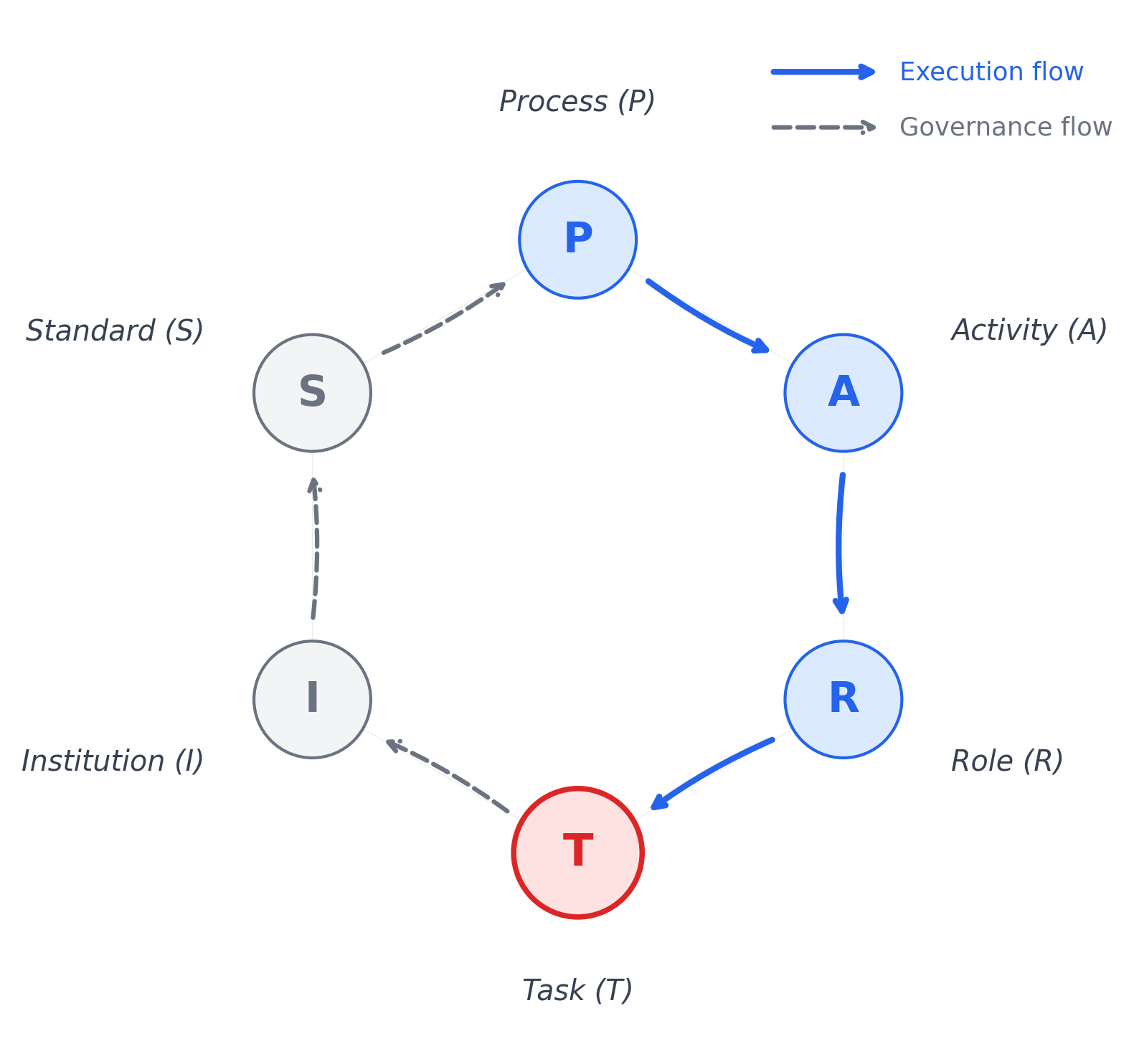}
\caption{PARTIS hexagonal architecture: six dimensions with the two flows (solid = Execution Flow; dashed = Governance Flow).}
\label{fig:1}
\end{figure}

\textbf{Execution Flow} (solid arrows): Process decomposes into Activities (P$\rightarrow$A), Activities are assigned to Roles (A$\rightarrow$R), Roles execute Tasks (R$\rightarrow$T). Each relationship is formalized as OCL constraints in the PARTIS metamodel (e.g., C1: every Process contains at least one Activity; C5: every Task has exactly one execution Role). \textbf{Governance Flow} (dashed arrows): Tasks are constrained by Institutions (T$\rightarrow$I), Institutions are codified into Standards (I$\rightarrow$S), Standards regulate Processes (S$\rightarrow$P). The pivot is \textbf{Task (T)}---simultaneously the Execution Flow terminus and the Governance Flow origin. \textbf{Institution (I)} captures regulatory constraints (e.g., securities law provisions); \textbf{Standard (S)} captures operational norms (e.g., CMMI level requirements). Their separation is empirically grounded: five senior architects unanimously confirmed that ``why the regulator requires this`` (I) and ``what specific standard must be followed`` (S) serve distinct decision needs.

\textbf{Governance cycle in practice.} Consider a task ``generate compliance report`` (T): securities law prohibits AI-generated investor-facing statements without human sign-off (I: regulatory constraint, T$\rightarrow$I). This institutional mandate is codified as Standard ``S-CR-04: all compliance reports require licensed-officer approval before release`` (I$\rightarrow$S). Standard S-CR-04 in turn constrains the parent process by requiring an approval gateway after the report-generation activity (S$\rightarrow$P). When the regulation changes---e.g., a new provision permits AI-generated internal reports without sign-off---the governance cycle propagates the change: I is updated, S-CR-04 is revised to scope the approval requirement to external reports only, and the process model is updated accordingly. This is not automated enforcement but a structured change-propagation protocol: each link in T$\rightarrow$I$\rightarrow$S$\rightarrow$P identifies which artifacts must be reviewed when an upstream element changes.

Fig. 2 illustrates the end-to-end PARTIS pipeline on a concrete example from the CM (Configuration Management) domain. The CM process decomposes into the CM.1 Code Management activity, which further decomposes into three tasks via T-IPO. Each task receives five-dimension LARA scores and a resulting level. CM.1.1 (Merge Request Review) scores L2 because D4 (Compliance Sensitivity) = 3 elevates the weighted mean above the L1 threshold---the governance constraint (securities law $\rightarrow$ Standard S-CR-04) requires human approval. CM.1.2 and CM.1.3 score L1, so full-agent deployment is possible. This single activity thus spans two LARA levels, confirming the intra-activity heterogeneity that motivates task-level assessment.

\begin{figure}[!htbp]
\centering
\includegraphics[width=0.737\linewidth,keepaspectratio]{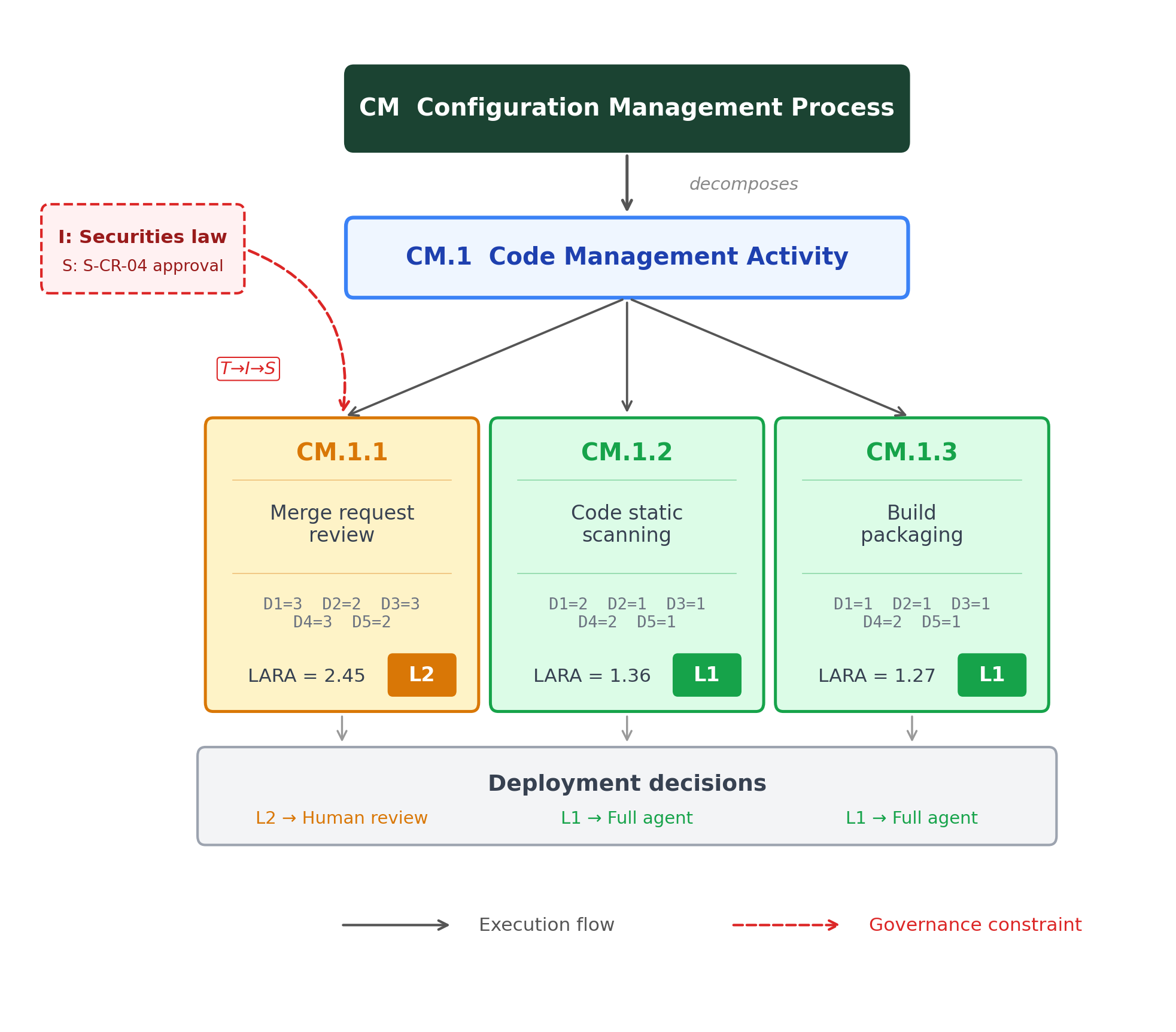}
\caption{End-to-end running example: CM.1 Code Management activity decomposed into three tasks with LARA scores, governance constraints (dashed), and deployment decisions.}
\label{fig:2}
\end{figure}

\section{The T-IPO Task Atomization Model}

\subsection{Design Rationale}

T-IPO addresses the granularity gap. In our dataset, the activity ``software architecture design`` decomposes into seven tasks spanning all four LARA levels. An activity-level assessment yields a misleadingly moderate score that masks both high-substitutability tasks (e.g., ``compile architecture document,`` L1) and irreducibly human tasks (e.g., ``architecture decision-making,`` L4). This intra-activity heterogeneity is empirically confirmed in \S6.4.

\subsection{Formal Definition}

\emph{\textbf{Definition 1 (Task Eight-Tuple).}} A task $\tau$ is defined as:

\emph{$\tau$ = (Id, Name, Role, Input, Logic, Output, Constraint, Dependency)}

where \emph{Id} is globally unique; \emph{Name} follows verb--noun convention; \emph{Role} $\in$ \{Human, LLMAgent, System, Hybrid\}, where System denotes deterministic software executors (e.g., scheduled jobs, RPA bots) requiring no cognitive assessment; \emph{Input} is a non-empty artifact set (C3); \emph{Logic} = (Steps, Tools, BloomLevel $\in$ \{1..6\}, Determinism $\in$ \{Deterministic, Probabilistic, Heuristic\}); \emph{Output} is a non-empty artifact set each with a Definition of Done (C4); \emph{Constraint} is a set of typed constraints (TimeConst, AuthConst, QualConst, AuditConst---e.g., audit trail retention $\geq$ 3 years) from the I and S dimensions; \emph{Dependency} is a set of typed directed pairs (Data, Temporal, Resource) forming a DAG (C6). BloomLevel takes the \emph{maximum} Bloom level across all Steps. Six key OCL constraints govern the metamodel (Table 2):

\begin{table}[!htbp]
\centering
\small
\renewcommand{\arraystretch}{1.1}
\begin{tabular}{@{}>{\raggedright\arraybackslash}p{0.0590\linewidth}
  >{\raggedright\arraybackslash}p{0.2361\linewidth}
  >{\raggedright\arraybackslash}p{0.6250\linewidth}@{}}
\toprule
\textbf{ID} & \textbf{Name} & \textbf{OCL expression (simplified)} \\
C1 & Process non-empty & context Process inv: self.activity$\rightarrow$size() $\geq$ 1 \\
C2 & Activity non-empty & context Activity inv: self.task$\rightarrow$size() $\geq$ 1 \\
C3 & Input non-empty & context Task inv: self.input$\rightarrow$size() $\geq$ 1 \\
C4 & Output non-empty + DoD & context Task inv: self.output$\rightarrow$size() $\geq$ 1 and self.output$\rightarrow$forAll(o \textbar{} o.dod$\rightarrow$notEmpty()) \\
C5 & Single execution role & context Task inv: self.executionRole$\rightarrow$size() = 1 \\
C6 & DAG acyclicity & context Task inv: self.allPredecessors()$\rightarrow$excludes(self) \\
\bottomrule
\end{tabular}
\caption{Key OCL constraints governing the T-IPO metamodel (simplified notation; full specification provided in Appendix A).}
\label{tab:2}
\end{table}

Fig. 3 expands the T-IPO structure of CM.1.2, one of the two L1 tasks identified in the running example (Fig. 2), showing how the eight-tuple elements map to a concrete task specification.

\begin{figure}[!htbp]
\centering
\includegraphics[width=0.737\linewidth,keepaspectratio]{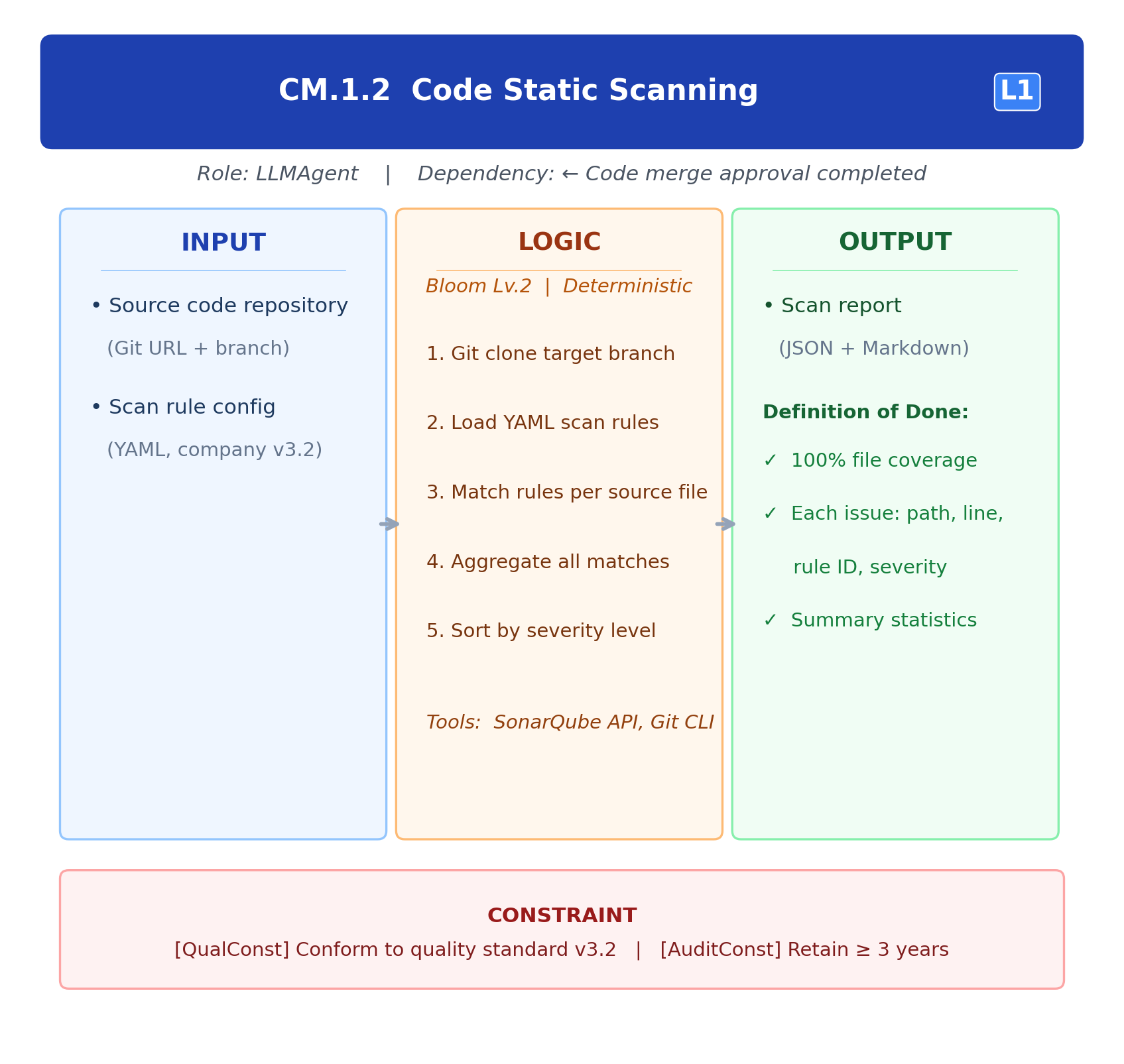}
\caption{T-IPO eight-tuple: worked example for CM.1.2 ``Code Static Scanning`` (L1, Bloom Lv.2, Deterministic).}
\label{fig:3}
\end{figure}

\subsection{Boundary Rules and Granularity Principles}

Four boundary rules: R1 (Single-Role): one execution role per task (OCL C5). R2 (Bloom Homogeneity): all steps at same or adjacent Bloom levels. R3 (Observable Output): every task produces at least one verifiable artifact with testable DoD. R4 (Minimal DAG): stop when further splitting creates tasks whose outputs are never independently consumed. Three granularity principles: P1 (Activity Anchoring), P2 (LARA Assessability), P3 (Agent Mappability)---the IPO triple must suffice to generate an unambiguous agent prompt.

\textbf{Decomposition reproducibility.} Two independent teams decomposed Configuration Management (6 activities) using T-IPO boundary rules. Task-set overlap: 85\% at semantic-equivalence level; Cohen`s $\kappa$ = 0.82 for LARA-level agreement on matched tasks---initial evidence that the rules yield consistent decompositions.

\subsection{T-IPO to Agent Prompt Mapping}

The eight-tuple maps to a four-section prompt architecture (Fig. 4): Role+Name $\rightarrow$ System Prompt; Input $\rightarrow$ Context; Logic $\rightarrow$ Instruction; Output+DoD $\rightarrow$ Format; Constraint $\rightarrow$ Guardrails.

\begin{figure}[!htbp]
\centering
\includegraphics[width=0.689\linewidth,keepaspectratio]{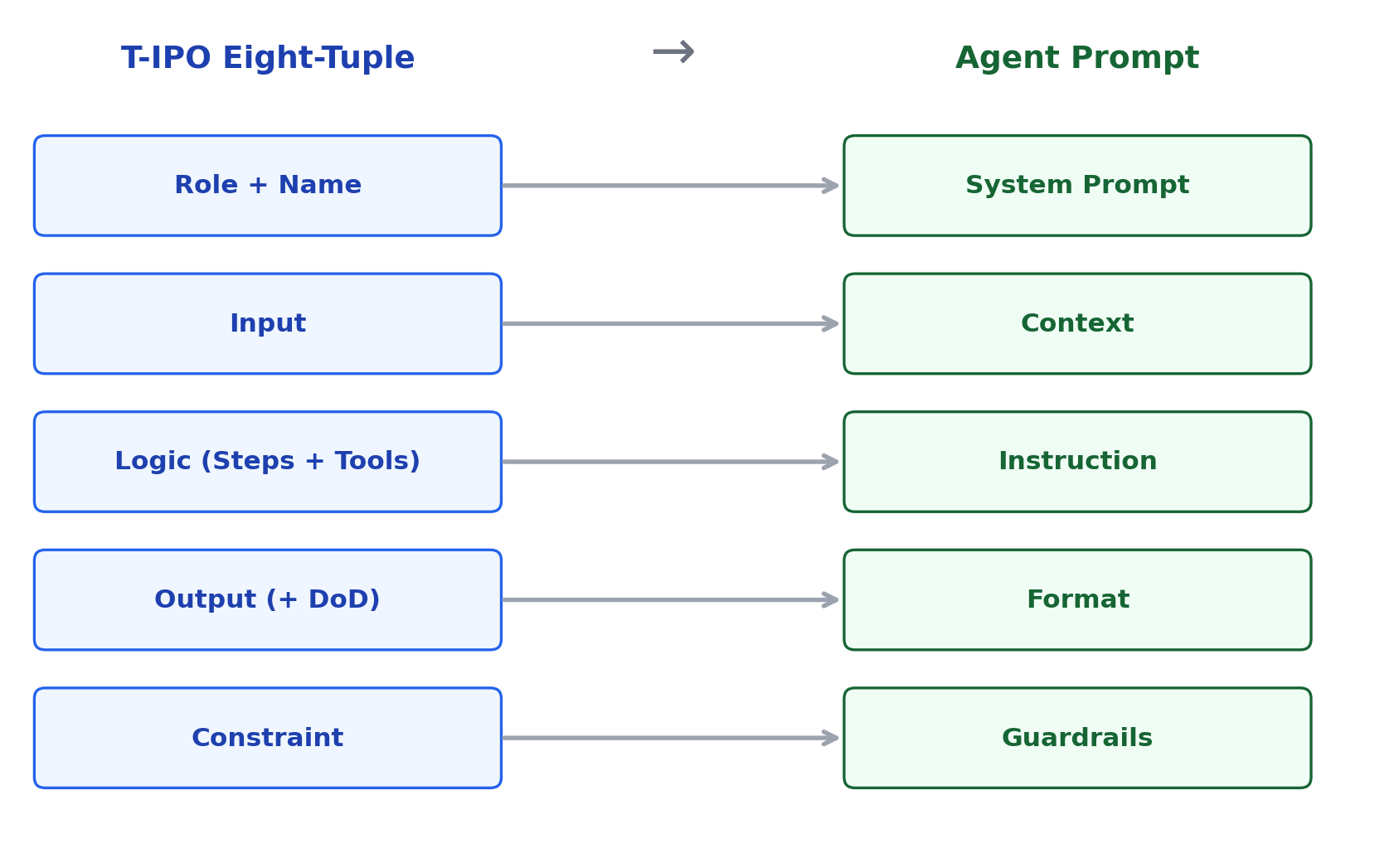}
\caption{Structural mapping from T-IPO eight-tuple to four-section agent prompt architecture.}
\label{fig:4}
\end{figure}

\textbf{Engineering impact.} In pilot deployment, we conducted a controlled before--after comparison on the same 6 L1 tasks (20 prompt pairs): switching from ad-hoc natural-language prompts to T-IPO-structured prompts improved auto-completion from 75\% to 95\%, reduced authoring time by \textasciitilde75\% (4 h $\rightarrow$ 1 h), and decreased cross-author output variability (CV from 0.60 to 0.15).

\section{The LARA Matrix}

\subsection{Dimension Derivation and Weight Validation}

LARA`s five dimensions were derived in three steps. Step 1 (Literature Synthesis): 14 factors from Autor et al. [7], Frey and Osborne [6], Brynjolfsson and Mitchell [1], Eloundou et al. [3], and securities-industry practice. Step 2 (Affinity Grouping): consolidated to five orthogonal dimensions (e.g., ``cognitive complexity,`` ``knowledge requirement,`` ``reasoning depth`` merged into D1 with Bloom anchors; ``regulatory compliance,`` ``audit traceability,`` ``explainability`` merged into D4). Step 3 (Delphi Validation): 12 experts, three rounds, Kendall`s W from 0.52 to 0.76. D4 received 27/100 points, which gave it a 1.5$\times$ weight.

\textbf{AHP cross-validation.} Eight Delphi experts completed AHP pairwise comparisons (Saaty`s [33] 1--9 scale). Aggregated weights: (0.190, 0.160, 0.180, 0.300, 0.170); mean CR = 0.021. All Delphi--AHP deviations \textless{} 0.03. D4 highest under both methods (Delphi: 0.273; AHP: 0.300 $\approx$ 1.5$\times$)---two independent methods converging on the same structure. Table 3 defines the resulting dimensions.

\subsection{Dimension Definitions}

\begin{table}[!htbp]
\centering
\small
\renewcommand{\arraystretch}{1.1}
\begin{tabular}{@{}>{\raggedright\arraybackslash}p{0.0605\linewidth}
  >{\raggedright\arraybackslash}p{0.1652\linewidth}
  >{\raggedright\arraybackslash}p{0.5841\linewidth}
  >{\raggedright\arraybackslash}p{0.1101\linewidth}@{}}
\toprule
\textbf{Dim.} & \textbf{Name} & \textbf{Definition and anchor} & \textbf{Weight} \\
D1 & Cognitive Complexity & Bloom-anchored: 1=Remember, 2=Understand, 3=Apply, 4=Analyze, 5=Evaluate/Create & 1.0$\times$ \\
D2 & Data Dependency & 1=single structured source $\rightarrow$ 5=real-time streaming + unstructured external data & 1.0$\times$ \\
D3 & Interaction Diversity & 1=fully independent $\rightarrow$ 5=cross-organizational political negotiation & 1.0$\times$ \\
D4 & Compliance Sensitivity & 1=no compliance $\rightarrow$ 5=maximum; errors may cause legal liability & \textbf{1.5$\times$} \\
D5 & Innovation Requirement & 1=fully templated $\rightarrow$ 5=breakthrough original solution required & 1.0$\times$ \\
\bottomrule
\end{tabular}
\caption{LARA five dimensions with scoring anchors and weights.}
\label{tab:3}
\end{table}

\textbf{Bloom as heuristic proxy.} D1 uses Bloom`s revised taxonomy [18] as an ordering heuristic for LLM capability, not as a theoretical isomorphism. LLMs exhibit a ``jagged`` capability profile [5]: GPT-4 can perform Bloom Lv.5 tasks on legal texts but fails at Bloom Lv.3 arithmetic. We collapse Bloom Lv.5 (Evaluate) and Lv.6 (Create) into D1=5 because both require generative judgment beyond current reliable LLM capacity in enterprise settings---a pragmatic simplification validated empirically ($\kappa$ = 0.80 confirms that raters can consistently apply the mapping) rather than theoretically guaranteed. This compression reduces the D1 scale from a 6-point to an effective 5-point range, slightly compressing discriminant power among the most cognitively demanding tasks---a limitation we accept given the small number of Lv.5--6 tasks in our dataset (N = 7, 5.5\% of 127). As LLM capabilities evolve, the Bloom--D1 calibration must be periodically updated via the LARA-TCA recalibration procedure (\S5.6).

\textbf{D4 weight as domain-specific calibration.} Our research setting is a heavily regulated industry, which likely raised the salience of D4 during our Delphi panel. The 1.5$\times$ D4 weight is a \emph{domain-specific parameter} derived from a Delphi panel predominantly composed of financial-services practitioners (9/12). Applying LARA outside financial services requires re-running the Delphi with industry-appropriate experts. For low-compliance industries (e.g., creative agencies), D4 weight may approach 1.0$\times$; for banking, Firm-B evaluators recommended 1.8--2.0$\times$. The W vector is explicitly designed to be recalibrable.

\subsection{Scoring, Thresholds, and Boundary Treatment}

The weighted mean is LARA($\tau$) = $\Sigma$(D$_i$$\times$W$_i$)/$\Sigma$W$_i$, with W = (1, 1, 1, 1.5, 1), $\Sigma$W = 5.5, range {[}1.0, 5.0{]}. Thresholds via k-means (k=4, silhouette 0.72) and Elbow: L1 {[}1.0, 2.0{]}; L2 (2.0, 3.0{]}; L3 (3.0, 4.0{]}; L4 (4.0, 5.0{]}. Sensitivity: $\pm$0.2 perturbations affect 5--6\% of tasks.

\textbf{Boundary-task treatment.} This was one of the first places we worried about judgment variance in early pilots. In the 127-task dataset, 11 tasks (8.7\%) fell within $\pm$0.15 of a threshold. Principle: ``conservative upgrade``---borderline tasks classified into the higher human-involvement level. For the L2/L3 boundary (2.85--3.15), D4 is the decisive swing: D4 $\geq$ 3 trends toward L3; D4 $\leq$ 2 toward L2. Organizations should maintain a ``boundary watch-list`` and re-evaluate after major model upgrades.

\textbf{Weight--threshold--migration mechanism.} Three parameter categories must be distinguished. \emph{Fixed parameters:} the five-dimension structure, the weighted-mean formula, and the four-level classification are design constants that define the method. \emph{Domain-recalibrable parameters:} the weight vector W and the Bloom--D1 anchor mapping require industry-specific Delphi when LARA is deployed in a new sector. \emph{Threshold-coupled parameters:} when W changes, thresholds must be re-estimated via k-means on the new industry`s task data, because the score distribution shifts. Consequently, L1--L4 labels are internally valid within a given industry calibration but \emph{not directly comparable across industries} with different W vectors---an L2 in banking (W$_4$ = 2.0) may correspond to a different absolute risk profile than an L2 in creative services (W$_4$ = 1.0). Cross-industry comparisons require normalizing to the equal-weight baseline before comparison.

\subsection{Four-Level Classification}

\begin{table}[!htbp]
\centering
\small
\renewcommand{\arraystretch}{1.1}
\begin{tabular}{@{}>{\raggedright\arraybackslash}p{0.0688\linewidth}
  >{\raggedright\arraybackslash}p{0.1082\linewidth}
  >{\raggedright\arraybackslash}p{0.2413\linewidth}
  >{\raggedright\arraybackslash}p{0.2558\linewidth}
  >{\raggedright\arraybackslash}p{0.2459\linewidth}@{}}
\toprule
\textbf{Level} & \textbf{Range} & \textbf{Verdict} & \textbf{Deployment mode} & \textbf{Human role} \\
L1 & 1.0--2.0 & High substitution & Full agent + 5\% spot check & 5\% audit \\
L2 & 2.1--3.0 & Assistive augmentation & Agent drafts + human approval & 100\% review \\
L3 & 3.1--4.0 & Selective assistance & Human-led + agent copilot & Human-led \\
L4 & 4.1--5.0 & Human-dominant & Fully human execution & 100\% human \\
\bottomrule
\end{tabular}
\caption{LARA four-level classification with deployment guidance.}
\label{tab:4}
\end{table}

Epistemologically, LARA is a deployment heuristic, not a deterministic classifier (Table 4). An L1 rating means ``a candidate worth pilot-validating under governance controls.`` When a pilot fails for an L1-rated task, the task is automatically reclassified to L2, and root-cause analysis is triggered to determine whether the failure reflects a scoring error, a technology limitation, or an edge-case data anomaly.

\subsection{Non-Compensable D4 Floor Rule}

The weighted-mean formula allows a task with D4 = 5 (maximum compliance sensitivity) but low scores on other dimensions to classify as L2 when they should be L3. In highly regulated settings, this compensability may underestimate governance risk. We therefore introduce a \emph{non-compensable floor constraint}:

\emph{\textbf{Definition 2 (D4 Floor Rule).} If D4 $\geq$ 4, then LARA level $\geq$ L3 regardless of the weighted-mean total. Formally: LARA$_s$(task) = max(LARA$_m$$_e$$_a$$_n$(task), L3) when D4(task) $\geq$ 4.}

This rule reflects the regulatory reality that high-compliance tasks require human oversight independent of cognitive simplicity. In our dataset, the floor affects 3 tasks (2.4\%) whose weighted mean places them at L2 but whose D4 = 4 elevates them to L3. The rule is optional and domain-configurable: financial-services deployments should enable it; low-regulation contexts (e.g., creative agencies) may disable it. The D4 threshold (4) and the floor level (L3) are themselves domain-recalibrable parameters within the weight--threshold--migration mechanism (\S5.3). Of the 7 tasks affected by Bloom Lv.5--6 compression (\S5.2), none has D4 $\geq$ 4---the intersection of high cognitive complexity and high compliance sensitivity is empty in our dataset. This suggests that the two parameter adjustments (Bloom compression, D4 floor) operate on disjoint task subsets, confirming their independence.

\subsection{The LARA-TCA Recalibration Procedure}

LARA scores reflect a snapshot of LLM capability at assessment time. As models evolve, scores require periodic recalibration. The LARA Technical Capability Assessment (LARA-TCA) procedure formalizes this: recalibration is triggered by (a) major model upgrades, (b) performance drift exceeding $\pm$0.5 on any dimension`s mean score (detected via quarterly monitoring), or (c) regulatory changes altering D4 anchors. The default cadence is six-monthly. The procedure comprises five steps: select a stratified sample ($\geq$20\% of tasks), re-administer LARA scoring with the current model, compute per-dimension drift ($\Delta$), trigger full reassessment if mean \textbar $\Delta$\textbar{} \textgreater{} 0.5 (threshold selected to exceed two standard errors of dimension-level scoring variability in our dataset, SE $\approx$ 0.22), and re-estimate thresholds via k-means if W-vector changes are warranted (\S5.3). Output: an updated scorecard and migration change log.

\textbf{Temporal stability.} We do not report test--retest reliability because the six-month recalibration window has not yet elapsed since the primary assessment (January--February 2025). Test--retest data collection is planned for Q3 2025. The absence of test--retest data is a limitation; LARA-TCA is the designed mitigation, positioning LARA as a dynamic governance instrument instead of a static classification.

\section{Empirical Validation}

\subsection{Study Design}

The empirical program follows a sequential mixed-methods design [48], comprising six components that together cover the main validity facets. Table 5 provides an overview.

\begin{table}[!htbp]
\centering
\small
\renewcommand{\arraystretch}{1.1}
\begin{tabular}{@{}>{\raggedright\arraybackslash}p{0.2164\linewidth}
  >{\raggedright\arraybackslash}p{0.1180\linewidth}
  >{\raggedright\arraybackslash}p{0.1770\linewidth}
  >{\raggedright\arraybackslash}p{0.1869\linewidth}
  >{\raggedright\arraybackslash}p{0.2217\linewidth}@{}}
\toprule
\textbf{Component} & \textbf{N} & \textbf{Data type} & \textbf{Validity facet} & \textbf{Primary metric} \\
Primary LARA study & 127 tasks $\times$ 3 raters & Quantitative & Inter-rater reliability & Fleiss` $\kappa$ = 0.80 \\
Cross-org. replication & 97 tasks $\times$ 3 orgs & Quantitative & External robustness & $\kappa$ = 0.73 \\
Comparison experiment & 49 tasks $\times$ 2 groups & Quantitative & Marginal value & +27--33 pp accuracy \\
Pilot deployment & 120 instances & Quant. + qual. & Criterion validity & L1 = 95\% auto \\
T-IPO reproducibility & 22 tasks $\times$ 2 teams & Quantitative & Decomp. stability & $\kappa$ = 0.82, 85\% overlap \\
Exploratory FA & 127 tasks & Quantitative & Structural validity & 2-factor, 64\% var. \\
\bottomrule
\end{tabular}
\caption{Multi-method validation program overview.}
\label{tab:5}
\end{table}

The primary study was conducted at Firm-A, a mid-sized Chinese securities firm (\textasciitilde2,000 employees; IT department \textasciitilde180). Five IT-operations domains: Project Management (PM: 8 activities, 32 tasks), Requirements Management (RM: 6/24), Software Architecture (SA: 5/21), Test Management (TM: 7/28), Configuration Management (CM: 6/22). T-IPO decomposition yielded 127 tasks (mean 3.9/activity, SD = 1.3). Three senior architects (8--15 years) served as independent raters after a one-day calibration workshop (initial $\kappa$ = 0.65; post-calibration $\kappa$ = 0.82; calibration excluded). Scoring: independent rating followed by structured dispute resolution. Statistical power analysis (Cohen 1988): at $\alpha$ = 0.05 and power = 0.80 with medium effect size (w = 0.30), the minimum sample for a four-category chi-square test (df = 3) is n = 122; our N = 127 satisfies this with 4\% margin.

\subsection{LARA Results (N = 127)}

\begin{table}[!htbp]
\centering
\small
\renewcommand{\arraystretch}{1.1}
\begin{tabular}{@{}>{\raggedright\arraybackslash}p{0.2069\linewidth}
  >{\raggedright\arraybackslash}p{0.0730\linewidth}
  >{\raggedright\arraybackslash}p{0.0827\linewidth}
  >{\raggedright\arraybackslash}p{0.0730\linewidth}
  >{\raggedright\arraybackslash}p{0.0730\linewidth}
  >{\raggedright\arraybackslash}p{0.0730\linewidth}
  >{\raggedright\arraybackslash}p{0.0730\linewidth}
  >{\raggedright\arraybackslash}p{0.0913\linewidth}
  >{\raggedright\arraybackslash}p{0.0827\linewidth}
  >{\raggedright\arraybackslash}p{0.0913\linewidth}@{}}
\toprule
\textbf{Domain} & \textbf{Act.} & \textbf{Tasks} & \textbf{L1} & \textbf{L2} & \textbf{L3} & \textbf{L4} & \textbf{L1\%} & \textbf{$\kappa$} & \textbf{D4$\mu$} \\
Project Mgmt & 8 & 32 & 16 & 10 & 4 & 2 & 50.0 & 0.78 & 2.3 \\
Requirements & 6 & 24 & 12 & 8 & 3 & 1 & 50.0 & 0.82 & 2.5 \\
Architecture & 5 & 21 & 7 & 8 & 4 & 2 & 33.3 & 0.74 & 2.8 \\
Test Mgmt & 7 & 28 & 16 & 8 & 3 & 1 & 57.1 & 0.85 & 2.0 \\
Config. Mgmt & 6 & 22 & 9 & 10 & 2 & 1 & 40.9 & 0.80 & 2.6 \\
\textbf{Total} & \textbf{32} & \textbf{127} & \textbf{60} & \textbf{44} & \textbf{16} & \textbf{7} & \textbf{47.2} & \textbf{0.80} & \textbf{2.4} \\
\bottomrule
\end{tabular}
\caption{LARA results (N = 127). $\kappa$ = Fleiss` kappa [14]; D4$\mu$ = mean Compliance Sensitivity. Results reflect pre-floor classification; applying Definition 2 reclassifies 3 L2 tasks to L3.}
\label{tab:6}
\end{table}

\begin{figure}[!htbp]
\centering
\includegraphics[width=0.737\linewidth,keepaspectratio]{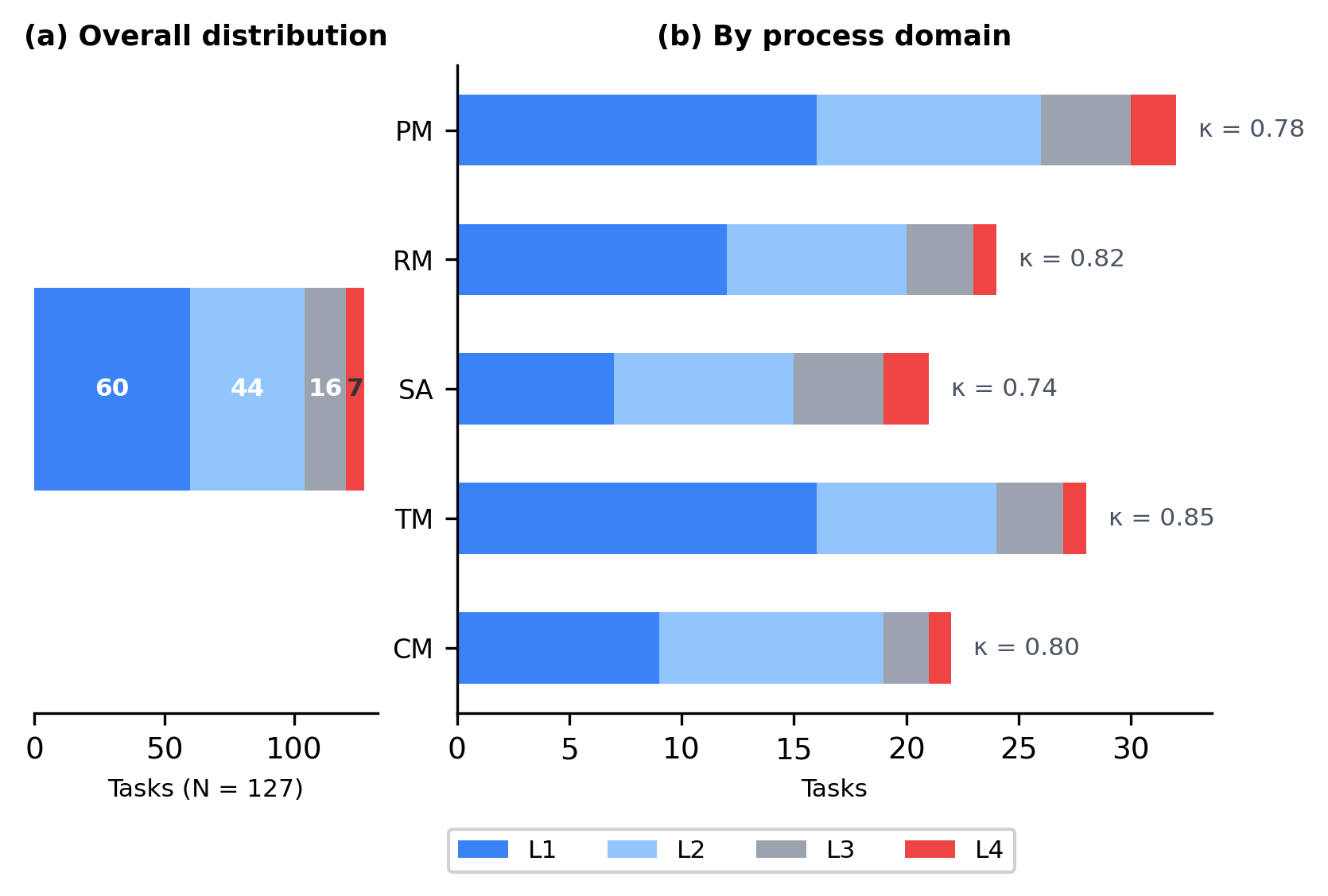}
\caption{LARA distribution: (a) overall; (b) by domain with $\kappa$ values.}
\label{fig:5}
\end{figure}

\textbf{Key findings (Table 6, Fig. 5).} (F1) Distribution: 47.2\% L1, 34.6\% L2, 12.6\% L3, 5.5\% L4. (F2) Reliability: Fleiss` $\kappa$ = 0.80 (Bootstrap 95\% CI {[}0.72, 0.87{]}); Gwet`s AC1 = 0.77 [15] (correcting for base-rate skew). Domain $\kappa$: 0.74 (SA) to 0.85 (TM). (F3) D4 effect: 1.5$\times$ weight shifted four compliance tasks from L1 to L2. (F4) Intra-activity heterogeneity: 78\% of activities span $\geq$2 LARA levels; the extreme case (``system architecture design``) spans L1--L4.

\subsection{Cross-Organizational Replication (N = 97)}

Three institutions independently replicated LARA: Firm-B (bank, 38 tasks, $\kappa$ = 0.72), Firm-C (insurance, 31 tasks, $\kappa$ = 0.75), Firm-D (securities fintech, 28 tasks, $\kappa$ = 0.78). All CI lower bounds \textgreater{} 0.60 (substantial per Landis and Koch [34]). Weighted mean $\kappa$ = 0.73. L1 proportion: 39.5--50.0\% (Kruskal--Wallis H = 3.42, p = 0.33).

\subsection{Comparison Experiment}

\emph{\textbf{Validity caveat.}} Three threats stand out. Group B received full LARA training while Group A had only a half-day generic briefing, so training asymmetry could be driving part of the gap. Group B's data may have been reused in subtle ways. The sample is also limited — two domains and 12 activities. The training asymmetry likely inflates Group B`s advantage; however, even under the conservative assumption that 50\% of the accuracy gap is attributable to training, the residual +13--17 pp difference remains practically significant. Moreover, in real organizational settings the training investment is part of the method cost: a practitioner choosing task-level assessment would necessarily invest in LARA training, making the ``unfair`` comparison ecologically realistic. The one-day LARA training represents less than 2\% of the total method cost (vs. $\sim$3 weeks for T-IPO decomposition of 127 tasks), further reducing the proportional significance of the training asymmetry. Conclusions are positioned as supportive evidence.

Two domains selected for heterogeneity contrast: SA (5 activities/21 tasks; L1=33\%) and TM (7/28; L1=57\%). Group A assessed activities holistically (1--5 scale, no T-IPO); Group B used LARA task scores aggregated. Validation: agent pilot (5 instances per L1 task, expert blind-evaluated). Fig. 6 illustrates the critical case and Table 7 quantifies the results.

\begin{figure}[!htbp]
\centering
\includegraphics[width=0.737\linewidth,keepaspectratio]{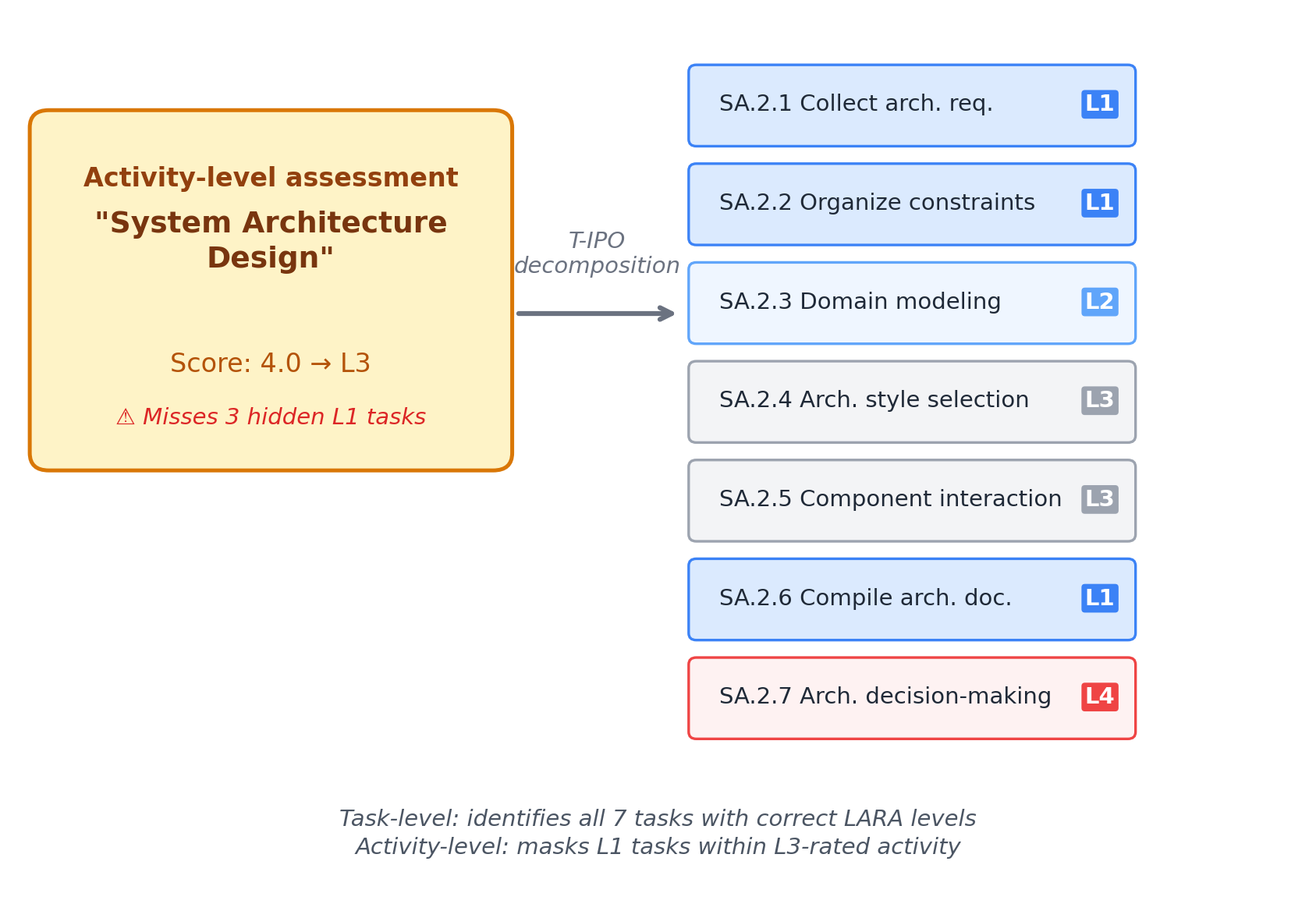}
\caption{Comparison: ``System Architecture Design`` receives L3 at activity level, masking three L1 tasks visible through T-IPO decomposition.}
\label{fig:6}
\end{figure}

\begin{table}[!htbp]
\centering
\small
\renewcommand{\arraystretch}{1.1}
\begin{tabular}{@{}>{\raggedright\arraybackslash}p{0.2655\linewidth}
  >{\raggedright\arraybackslash}p{0.2065\linewidth}
  >{\raggedright\arraybackslash}p{0.2065\linewidth}
  >{\raggedright\arraybackslash}p{0.2413\linewidth}@{}}
\toprule
\textbf{Metric} & \textbf{Group A (activity)} & \textbf{Group B (task)} & \textbf{Difference} \\
Prediction accuracy (SA) & 67\% & 100\% & +33 pp \\
Prediction accuracy (TM) & 73\% & 100\% & +27 pp \\
Missed-opportunity rate (SA) & 60\% & 0\% & -60 pp \\
Missed-opportunity rate (TM) & 33\% & 0\% & -33 pp \\
\bottomrule
\end{tabular}
\caption{Task-level vs. activity-level assessment.}
\label{tab:7}
\end{table}

A fair comparison must also consider cost: T-IPO decomposition required approximately 3 weeks of collaborative effort for 127 tasks, whereas activity-level assessment is near-instantaneous. The accuracy-per-effort trade-off favors T-IPO when the downstream cost of missed L1 opportunities (unrealized automation savings) or false-positive deployments (compliance incidents) is high---precisely the conditions prevailing in regulated industries.

\textbf{Note on 100\% Group B accuracy.} Group B`s perfect prediction accuracy reflects a ceiling effect of the small sample (12 activities): with only 2 domains and full LARA training, the task-level decomposition left no ambiguity about which tasks were L1. This ceiling effect interacts with the training asymmetry---even a small training-induced boost suffices to produce 100\% in a 12-activity sample, though the training itself represents less than 2\% of total method cost. In a larger sample spanning more heterogeneous domains, some misclassifications are expected---particularly at the L2/L3 boundary where D4 is the decisive swing dimension. The 100\% figure should therefore be interpreted as ``no errors in this small, well-characterized sample`` --- not ``the method is infallible.``

\subsection{Pilot Deployment (N = 120)}

Six L1 tasks (2 per domain) piloted in sandbox (20 instances each). L1 auto-completion: 95\% (114/120). L2: \textasciitilde70\% draft-quality. L3: \textasciitilde40\% useful suggestions. Cochran--Armitage trend: p \textless{} 0.001.

\textbf{Failure case analysis.} The 6 failures (5\%) clustered in two root causes. (a) \emph{Legacy-system data anomaly} (4/6): task CM.1.3 (``Build Packaging``, LARA = 1.36) failed on instances involving a legacy subsystem that produced non-standard CSV encoding (GBK with mixed line endings); the agent parsed the file incorrectly, generating an incomplete build manifest. The LARA score was correct---the task \emph{is} cognitively simple---but the input data deviated from the T-IPO specification`s assumed format. This failure reveals a gap in T-IPO`s Input element: format specifications should include tolerance ranges for encoding variants, not just the canonical format. (b) \emph{Edge-case DoD violation} (2/6): task TM.3.5 (``Test Report Generation``, LARA = 1.55) produced reports that met all DoD criteria except one instance where a zero-defect module was omitted from the summary table. The agent read ``zero defects`` as ``no entry needed`` when the specification actually required an explicit zero-row. This was resolved by refining the DoD to specify ``all modules listed regardless of defect count.`` Both failure types confirm LARA`s epistemological positioning: L1 is a deployment \emph{candidate}, not a deployment guarantee. Pilot validation remains essential. Methodologically, these pilot failures directly inform refinement of T-IPO`s specification rules: the Input element now requires encoding-tolerance ranges (not just canonical formats), and the DoD element requires explicit handling of zero-count edge cases---closing the DSR build--evaluate--refine loop.

\subsection{Exploratory Factor Analysis}

EFA on the polychoric correlation matrix [35] (N = 127, KMO = 0.72, Bartlett p \textless{} 0.001) using principal-axis factoring with oblimin rotation revealed a two-factor structure (64\% variance explained; F1--F2 correlation r = 0.24). F1 (Cognitive-Task Complexity) loads primarily on D1 (0.72), D5 (0.65), D3 (0.56), and D2 (0.44); F2 (Compliance Constraint) loads on D4 (0.78). All inter-dimension Spearman correlations \textless{} 0.50; VIF range 1.13--1.38 [36]. The two-factor structure is structurally consistent with---but does not confirm---the dual-cycle architecture: F1 aligns with the Execution Flow; F2 with the Governance Flow. We advance this as a testable proposition: task-level AI readiness is jointly determined by cognitive-execution complexity and governance-compliance intensity, and these two factors are empirically separable. The present EFA provides preliminary support; CFA on an independent cross-industry sample (N \textgreater{} 300) would constitute a stronger test. We retain five dimensions for: (a) operational feasibility (each has concrete anchors); (b) diagnostic value (same-total tasks with different D1/D4 profiles need different strategies); (c) weight customizability.

\subsection{Validation Boundaries and Extension Roadmap}

Three validation streams are planned to address the current study`s boundaries (Fig. 7). The CFA stream will test two competing models (two-factor vs. five-factor) on N \textgreater{} 300 tasks from 3+ industries, with fit evaluated via CFI $\geq$ 0.95, RMSEA $\leq$ 0.06, and SRMR $\leq$ 0.08 [43]; multi-group CFA will test cross-industry measurement invariance---failure would falsify Proposition 1 for the general case. The L3/L4 stream addresses the gap that LARA`s predictive validity is confirmed only for L1 tasks (95\% pilot), via expert-panel judgment (N = 10, 30 tasks) and longitudinal deployment tracking at Firm-A. The multi-domain stream will replicate T-IPO decomposition ($\kappa$ = 0.82 on one domain) across all five domains with independent team pairs.

\begin{figure}[!htbp]
\centering
\includegraphics[width=0.689\linewidth,keepaspectratio]{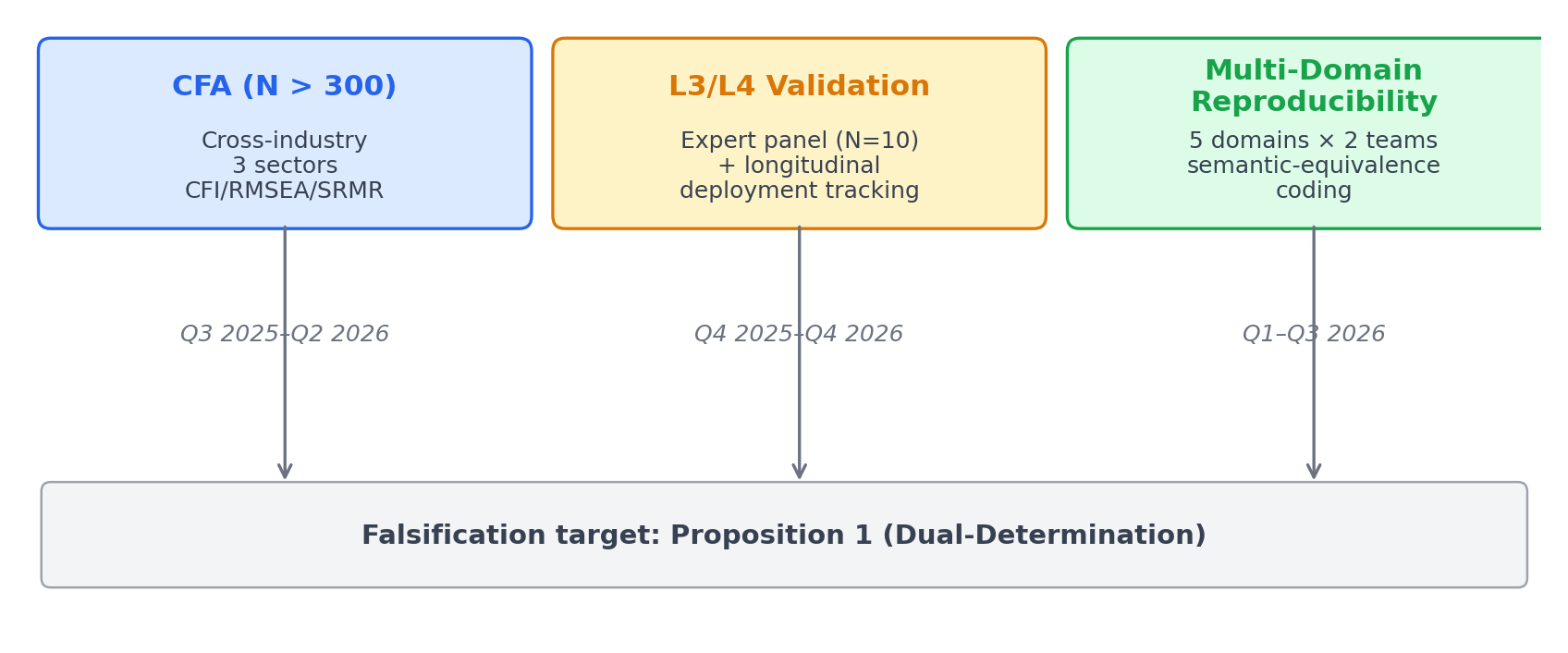}
\caption{Validation extension roadmap: three parallel streams targeting structural validity (CFA), criterion validity (L3/L4), and decomposition stability (multi-domain T-IPO reproducibility).}
\label{fig:7}
\end{figure}

\section{Discussion}

\subsection{Contributions}

The paper`s contributions are organized in three layers. At the \emph{framing level}, PARTIS provides a six-dimensional analytical architecture with dual execution--governance cycles, positioning Task as the pivot between ``how work is done`` and ``how work is governed.`` PARTIS is not itself the primary artifact but the \emph{analytical lens} that motivates and situates the two core artifacts. At the \emph{artifact level}, T-IPO (task-level structuring as eight-tuples with OCL constraints and boundary rules) and LARA (five-dimension readiness scoring built on Bloom's taxonomy with Delphi--AHP validated weights) constitute the paper`s primary design-science contributions, offering respectively a \emph{modeling primitive} (T-IPO pushes BPM granularity below the activity level) and a \emph{decision instrument} (LARA maps assessment scores to actionable deployment modes). At the \emph{evidence level}, the multi-method evaluation provides initial empirical support for these artifacts` inter-rater reliability, cross-organizational robustness, pilot utility, and preliminary predictive usefulness within the target context of financial-services IT operations.

\subsection{DSR Evaluation}

Table 8 maps our artifacts against Hevner et al.`s [10] seven guidelines. Per Venable et al.`s [37] FEDS: primarily summative-naturalistic (real-world deployment) with summative-artificial elements (comparison experiment) alongside.

\begin{table}[!htbp]
\centering
\small
\renewcommand{\arraystretch}{1.1}
\begin{tabular}{@{}>{\raggedright\arraybackslash}p{0.0787\linewidth}
  >{\raggedright\arraybackslash}p{0.2164\linewidth}
  >{\raggedright\arraybackslash}p{0.6250\linewidth}@{}}
\toprule
\textbf{G\#} & \textbf{Guideline} & \textbf{Evidence in this paper} \\
G1 & Design artifact & T-IPO (Def. 1), LARA (\S5), D4 Floor (Def. 2), LARA-TCA (\S5.6) \\
G2 & Problem relevance & Four practitioner-motivated gaps (\S1); industry validation at Firm-A \\
G3 & Design evaluation & Multi-method: $\kappa$, comparison experiment, pilot, EFA, cross-org replication (Table 5) \\
G4 & Research contribution & Level 2 nascent design theory [13]; Proposition 1 (\S7.3) \\
G5 & Research rigor & OCL constraints, Delphi--AHP dual validation, statistical power analysis \\
G6 & Design as search & 3 iterations, 3 alternatives eliminated (below); design log maintained \\
G7 & Communication & This paper (research) + deployment guidance (practice, \S7.4) \\
\bottomrule
\end{tabular}
\caption{Mapping against Hevner et al.`s [10] seven DSR guidelines.}
\label{tab:8}
\end{table}

\textbf{Design iterations and alternatives eliminated.} Three major design alternatives were considered and discarded. (1) A three-dimension LARA variant (D1/D4/D5 only) achieved lower $\kappa$ = 0.61 because D2 and D3 captured variance that D1 alone could not absorb. (2) An equal-weight design misclassified 4 compliance-sensitive tasks as L1, unanimously rejected by experts. (3) A continuous-score LARA was abandoned because practitioners reported that continuous scores lacked actionable thresholds---``tell me L1 or L2, not 2.37`` (Firm-A architect). Each elimination was documented and justified by empirical evidence.

\subsection{Toward a BPM Theory Proposition}

The EFA`s two-factor structure (\S6.6) motivates a nascent theoretical proposition for the BPM community:

\begin{table}[!htbp]
\centering
\small
\renewcommand{\arraystretch}{1.1}
\begin{tabular}{@{}>{\raggedright\arraybackslash}p{0.9200\linewidth}@{}}
\toprule
\emph{\textbf{Proposition 1 (Dual-Determination).} Task-level AI readiness in enterprise workflows is jointly determined by two empirically separable constructs: (a) cognitive-execution complexity (the aggregate of cognitive demand, data requirements, interaction patterns, and innovation needs) and (b) governance-compliance intensity (the degree to which regulatory, audit, and institutional constraints restrict autonomous agent execution).} \\
\bottomrule
\end{tabular}
\end{table}

This proposition is falsifiable: if CFA on cross-industry data yields a single-factor solution, or if the two-factor structure fails measurement invariance across industries, the proposition is rejected. The practical implication is that organizations cannot assess AI readiness by evaluating only task complexity (the execution dimension) or only compliance risk (the governance dimension)---both are necessary and empirically distinct determinants. We read this dual-determination structure directly reflects PARTIS`s architectural choice of dual cycles and provides a theoretical justification for why both cycles are needed.

\subsection{Practical Implications and Cross-Industry Applicability}

LARA`s four levels map directly to deployment modes with governance requirements. T-IPO-to-prompt mapping transforms prompt engineering from ``creative writing`` to ``structured filling.`` The boundary-task treatment operationalizes precautionary governance.

\textbf{Cross-industry applicability sketch.} While empirical validation is limited to financial-services IT operations, we illustrate how LARA would be recalibrated for a contrasting context. In a \emph{creative-agency} setting (e.g., advertising content production), compliance sensitivity is minimal---D4 weight would approach 1.0$\times$ (equal weight), because regulatory constraints rarely govern creative output. Conversely, D5 (Innovation Requirement) becomes the critical dimension, as creative work`s core value lies in novelty. A hypothetical creative-agency Delphi might yield W = (1, 1, 1, 1, 1.5), effectively inverting the financial-services weighting. The L1--L4 thresholds would then be re-estimated via k-means on creative-task data, likely with a lower L1 ceiling (more tasks classified as requiring human creativity). This sketch demonstrates that the LARA \emph{method} (five dimensions, weighted mean, four levels) is portable across industries, while the \emph{parameters} (W vector, thresholds, Bloom anchors) require domain-specific calibration---a distinction between method-level and parameter-level validity. Notably, in creative industries the Bloom anchors themselves may need restructuring, not just reweighting: Bloom Lv.6 (Create) tasks are rare outliers in financial IT but constitute the industry`s core value proposition in advertising, requiring qualitatively different D1 anchor descriptions. We emphasize that this cross-industry sketch is \emph{illustrative — not empirically validated}; actual deployment in non-financial domains requires a full Delphi--AHP recalibration cycle with domain-appropriate experts.

\subsection{Threats to Validity}

\textbf{Construct validity.} D5 exhibited weakest independence in EFA (loading on F1 with D1/D3); external criterion measures (e.g., Kirton`s KAI scale) are needed. The Bloom--D1 mapping is a heuristic proxy validated for the current LLM generation ($\kappa$ = 0.80) but requiring recalibration as models evolve (\S5.6). \textbf{Internal validity.} The evidence chain contains a bootstrapping structure (T-IPO defines tasks $\rightarrow$ LARA evaluates $\rightarrow$ pilot validates using T-IPO-defined tasks). LARA`s predictive validity is confirmed only for L1; L3/L4 external validation is planned (\S6.7, Fig. 7). The comparison experiment`s threats are detailed in \S6.4. \textbf{External validity.} All data from Chinese financial-services IT operations; cross-industry generalization requires D4 recalibration and Bloom-anchor adaptation (\S7.4). \textbf{Conclusion validity.} EFA is exploratory (N = 127); CFA on independent N \textgreater{} 300 is planned (\S6.7). Pilot sample (120 instances, 6 tasks) and single-domain reproducibility ($\kappa$ = 0.82) limit generalizability; multi-domain replication is planned (Fig. 7).

\section{Conclusion and Future Work}

This paper introduced T-IPO and LARA, a pair of DSR artifacts that work together for task-level AI readiness assessment, part of the larger PARTIS methodology (itself organized around two interacting cycles). T-IPO pushes BPM modeling granularity below the activity level by formalizing tasks as eight-tuples with OCL constraints and boundary rules, producing specifications directly mappable to LLM-agent prompt architectures. LARA is a five-dimension scoring matrix built on Bloom's taxonomy with Delphi--AHP validated weights. It produces a four-level classification that maps to actionable deployment modes with governance requirements.

Relative to the shortened workshop version (see footnote 1), this complete preprint adds several substantive pieces. The non-compensable D4 floor rule (Definition 2, \S5.5) is a new safeguard: it prevents the weighted mean from masking high governance risk when D4 is extreme. \S5.6 introduces LARA-TCA, a recalibration procedure that turns LARA from a static rubric into a dynamic instrument that tracks model progress. We develop the BWW ontological mapping in \S2.7, something the workshop version only hinted at. \S7.3 formalizes Proposition 1, a testable statement that positions the two-cycle architecture as a claim BPM theory can engage with. Finally, \S6.7 lays out a validation roadmap covering CFA design, L3/L4 external validation, and multi-domain reproducibility.

Empirical validation across 127 tasks demonstrated substantial reliability ($\kappa$ = 0.80), cross-organizational robustness ($\kappa$ = 0.73, 3 institutions), preliminary marginal-value evidence (+27--33 pp accuracy improvement, with training-asymmetry caveats), pilot criterion validity (L1 = 95\% auto-completion, monotonic L1$\rightarrow$L3 decline, p \textless{} 0.001), and a dual-factor EFA structure consistent with Proposition 1.

\textbf{Broader implications for the BPM community.} The shift from activity-level to task-level assessment is not merely a granularity refinement---it changes the class of questions BPM models can answer. Traditional process models answer ``what activities comprise this process and in what order?`` PARTIS-augmented models additionally answer ``which tasks within each activity can an LLM agent execute, under what governance constraints, and with what deployment mode?`` This capability is increasingly urgent as organizations move from RPA (deterministic, rule-based) to cognitive automation (probabilistic, LLM-based) [41]. The non-compensable D4 floor rule (\S5.5) operationalizes the principle that in regulated industries, governance is not a dimension to be traded off against efficiency but a hard constraint that bounds the automation design space.

Future directions: (1) CFA with cross-industry samples (N \textgreater{} 300, \S6.7) to test Proposition 1. (2) Controlled comparison experiment with balanced training to resolve the training-asymmetry threat. (3) L3/L4 external validation via expert-panel judgment and longitudinal deployment tracking. (4) Multi-domain decomposition reproducibility for T-IPO boundary rules across all five domains. (5) Multi-agent modeling: extending the Role dimension to sub-typed agent roles (planner, executor, reviewer) reflecting emerging multi-agent orchestration patterns [2, 23, 44]. (6) PARTIS Workbench: a software tool implementing T-IPO decomposition, LARA scoring, and LARA-TCA recalibration as an integrated workflow.


\begin{thebibliography}{99}
\footnotesize
\setlength{\itemsep}{1pt plus 0.2ex}
\setlength{\parsep}{0pt}
\bibitem{ref1} Brynjolfsson, E., Mitchell, T.: What can machine learning do? Workforce implications. Science 358(6370), 1530--1534 (2017)

\bibitem{ref2} OpenAI: GPT-4 technical report. arXiv:2303.08774 (2023)

\bibitem{ref3} Eloundou, T., Manning, S., Mishkin, P., Rock, D.: GPTs are GPTs: Labor market impact potential of LLMs. Science 384(6702), 1306--1308 (2024)

\bibitem{ref4} OMG: Business Process Model and Notation (BPMN), Version 2.0 (2011)

\bibitem{ref5} Dell`Acqua, F., et al.: Navigating the jagged technological frontier. Harvard Business School WP No. 24-013 (2023)

\bibitem{ref6} Frey, C.B., Osborne, M.A.: The future of employment. Technological Forecasting and Social Change 114, 254--280 (2017)

\bibitem{ref7} Autor, D.H., Levy, F., Murnane, R.J.: The skill content of recent technological change. QJE 118(4), 1279--1333 (2003)

\bibitem{ref8} Rogers, E.M.: Diffusion of Innovations. 5th edn. Free Press (2003)

\bibitem{ref9} Kotter, J.P.: Leading Change. Revised edn. Harvard Business Review Press (2012)

\bibitem{ref10} Hevner, A.R., et al.: Design science in IS research. MIS Quarterly 28(1), 75--105 (2004)

\bibitem{ref11} Simon, H.A.: The Sciences of the Artificial. 3rd edn. MIT Press (1996)

\bibitem{ref12} Peffers, K., et al.: A DSR methodology for IS research. JMIS 24(3), 45--77 (2007)

\bibitem{ref13} Gregor, S., Hevner, A.R.: Positioning and presenting DSR for maximum impact. MIS Quarterly 37(2), 337--355 (2013)

\bibitem{ref14} Fleiss, J.L.: Measuring nominal scale agreement among many raters. Psychological Bulletin 76(5), 378--382 (1971)

\bibitem{ref15} Gwet, K.L.: Handbook of Inter-Rater Reliability. 4th edn. Advanced Analytics (2014)

\bibitem{ref16} Wei, J., et al.: Chain-of-thought prompting elicits reasoning in LLMs. In: NeurIPS 2022

\bibitem{ref17} Arntz, M., Gregory, T., Zierahn, U.: The risk of automation for jobs in OECD countries. OECD WP No. 189 (2016)

\bibitem{ref18} Anderson, L.W., Krathwohl, D.R.: A Taxonomy for Learning, Teaching, and Assessing. Longman (2001)

\bibitem{ref19} Dumas, M., et al.: AI-augmented BPM systems: A research manifesto. ACM TMIS 14(1), Art. 11 (2023)

\bibitem{ref20} Kaltenpoth, S., Skolik, A., M\"uller, O., Beverungen, D.: A step towards cognitive automation: Integrating LLM agents with process rules. In: BPM 2025, LNCS, vol. 16044. Springer (2026)

\bibitem{ref21} Weske, M.: Business Process Management. 3rd edn. Springer (2019)

\bibitem{ref22} van der Aalst, W.M.P.: Process Mining. 2nd edn. Springer (2016)

\bibitem{ref23} Hughes, L., Dwivedi, Y.K., Malik, T., et al.: AI agents and agentic systems: A multi-expert analysis. Journal of Computer Information Systems 65(4), 1--29 (2025)

\bibitem{ref24} Stanton, N.A.: Hierarchical task analysis. Applied Ergonomics 37(1), 55--79 (2006)

\bibitem{ref25} Card, S.K., Moran, T.P., Newell, A.: The Psychology of Human--Computer Interaction. Erlbaum (1983)

\bibitem{ref26} Crandall, B., Klein, G., Hoffman, R.R.: Working Minds. MIT Press (2006)

\bibitem{ref27} European Parliament: Regulation (EU) 2024/1689 -- AI Act (2024)

\bibitem{ref28} NIST: AI Risk Management Framework (AI RMF 1.0). Gaithersburg (2024)

\bibitem{ref29} Floridi, L., et al.: AI4People. Minds and Machines 28(4), 689--707 (2018)

\bibitem{ref30} Scott, W.R.: Institutions and Organizations. 4th edn. Sage (2014)

\bibitem{ref31} Zachman, J.A.: A framework for IS architecture. IBM Systems Journal 26(3), 276--292 (1987)

\bibitem{ref32} Wand, Y., Weber, R.: On the deep structure of IS. ISJ 5(3), 203--223 (1995)

\bibitem{ref33} Saaty, T.L.: The Analytic Hierarchy Process. McGraw-Hill (1980)

\bibitem{ref34} Landis, J.R., Koch, G.G.: Observer agreement for categorical data. Biometrics 33(1), 159--174 (1977)

\bibitem{ref35} Holgado-Tello, F.P., et al.: Polychoric vs Pearson correlations in EFA. Quality \& Quantity 44(1), 153--166 (2010)

\bibitem{ref36} Hair, J.F., et al.: Multivariate Data Analysis. 8th edn. Cengage (2019)

\bibitem{ref37} Venable, J., et al.: FEDS: A framework for evaluation in DSR. EJIS 25(1), 77--89 (2016)

\bibitem{ref38} Noy, S., Zhang, W.: Experimental evidence on the productivity effects of generative AI. Science 381(6654), 187--192 (2023)

\bibitem{ref39} Ly, L.T., et al.: A framework for the systematic comparison and evaluation of compliance monitoring approaches. In: EDOC 2015. IEEE, pp. 7--16 (2015)

\bibitem{ref40} OMG: Decision Model and Notation (DMN), Version 1.1 (2016)

\bibitem{ref41} van der Aalst, W.M.P., et al.: Robotic process automation. BISE 60(4), 269--272 (2018)

\bibitem{ref42} Acemoglu, D., Restrepo, P.: Automation and new tasks: How technology displaces and reinstates labor. JEP 33(2), 3--30 (2019)

\bibitem{ref43} Hu, L., Bentler, P.M.: Cutoff criteria for fit indexes. Structural Equation Modeling 6(1), 1--55 (1999)

\bibitem{ref44} Hong, S., et al.: MetaGPT: Meta programming for multi-agent collaboration. In: ICLR 2024

\bibitem{ref45} Wand, Y., Weber, R.: On the ontological expressiveness of IS analysis and design grammars. ISJ 3(4), 217--237 (1993)

\bibitem{ref46} North, D.C.: Institutions, Institutional Change and Economic Performance. Cambridge UP (1990)

\bibitem{ref47} Wang, H., Poskitt, C.M., Sun, J.: AgentSpec: Customizable runtime enforcement for safe and reliable LLM agents. In: ICSE 2026. ACM (2026)

\bibitem{ref48} Creswell, J.W., Creswell, J.D.: Research Design. 5th edn. Sage (2018)

\bibitem{ref49} McKinsey Global Institute: The economic potential of generative AI. McKinsey (2023)

\bibitem{ref50} Reichert, M., Weber, B.: Enabling Flexibility in Process-Aware Information Systems. Springer (2012)
\end{thebibliography}
\end{document}